\documentclass[a4paper,11pt]{article}

\usepackage{pos}
\usepackage{caption}
\usepackage{subcaption}
\usepackage[table]{xcolor}
\usepackage{multirow}
\usepackage{array}
\usepackage{booktabs}
\usepackage{wrapfig}
\usepackage{enumitem}

\title{Comprehensive Study of the Lunar Energetic Particle Environment with LunPAN}
\author*[a]{J. Hulsman}
\author[e]{E. Alessi}
\author[f]{L. Andreasi}
\author[a]{P. Azzarello}
\author[b]{B. Bergmann}
\author[d]{M. Bolis}
\author[c]{P. Burian}
\author[a]{F. Cadoux}
\author[g]{M. Campbell}
\author[d]{C. Colombo}
\author[b]{H. Cintas}
\author[f]{A. Delille}
\author[b,a]{J. Jelinek}
\author[g]{X. LLorpart}
\author[f]{H. Santos}
\author[f]{M. Viitala}
\author[a]{X. Wu}

\affiliation[a]{Département de Physique Nucléaire et Corpusculaire (DPNC), University of Geneva, 24 Quai Ernest-Ansermet, Geneva, Switzerland}
\affiliation[b]{Institute for Experimental and Applied Physics (IEAP), Czech Technical University in Prague, Husova 240/5, 11000 Prague,Czech Republic}
\affiliation[c]{Faculty of Electrical Engineering, University of West Bohemia, Univerzitni 26, Pilsen, Czech Republic}
\affiliation[d]{Dipartimento di Scienze e Tecnologie Aerospaziale (DAER), Politecnico di Milano, Via La Masa, 34, Milano, 20156, Italy}
\affiliation[e]{Istituto di Matematica Applicata e Tecnologie Informatiche, Consiglio Nazionale delle Ricerche, Via Alfonso Corti 12, Milano, 20133, Italy}
\affiliation[f]{Aerospacelab S.A., Rue André Dumont 14b, 1435 Mont-Saint-Guibert, Belgium}
\affiliation[g]{European Organization for Nuclear Research (CERN) Esplanadedes Particules 1, 1211 Geneva 23, Switzerland}

\emailAdd{johannes.hulsman@unige.ch}

\abstract{LunPAN (Lunar Particle Analyzer Network) is a three-year mission proposal designed to comprehensively map the particle spectra in the lunar radiation field. It aims to provide precise measurements of Galactic Cosmic Rays (GCR), Solar Energetic Particles (SEP), and albedo particles, including charged particles, neutrons, and gamma-rays, originating from the Moon's surface. Therefore it will contribute to fundamental space physics, lunar geology sciences, space weather prediction, and radiation risk assessment for future lunar explorations. This is achieved through two state-of-the-art instruments; Pix.PAN and NeuPix. Pix.PAN is a compact magnetic spectrometer designed for precise measurements of penetrating charged particles, ranging from 100 MeV to 10 GeV. Based on the Mini.PAN project, Pix.PAN employs thin silicon pixel sensors optimized for energy resolution and particle identification. NeuPix is a hybrid active pixel sensor system capable of detecting neutrons, gamma-rays, and lower-energy charged particles between 10MeV and 100 MeV. Utilizing innovative sensor-converter combinations, NeuPix will provide spectral measurements of lunar albedo neutrons and gamma-ray fluxes. Currently, the LunPAN mission is accepted by ESA’s “Small Missions for Exploration – Destination the Moon” call for a pre-A phase study. We will discuss mission outline and expected scientific performance of the PixPAN and NeuPix.}

\ConferenceLogo{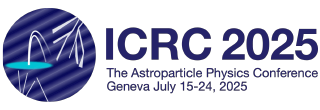}

\FullConference{39th International Cosmic Ray Conference (ICRC2025)\\
 15–24 July 2025\\
Geneva, Switzerland\\}
\begin{document}
\maketitle

\section{Introduction}

\begin{figure}[ht]
     \centering
     \begin{subfigure}[b]{0.45\textwidth}
         \centering
         \includegraphics[width=0.5\textwidth]{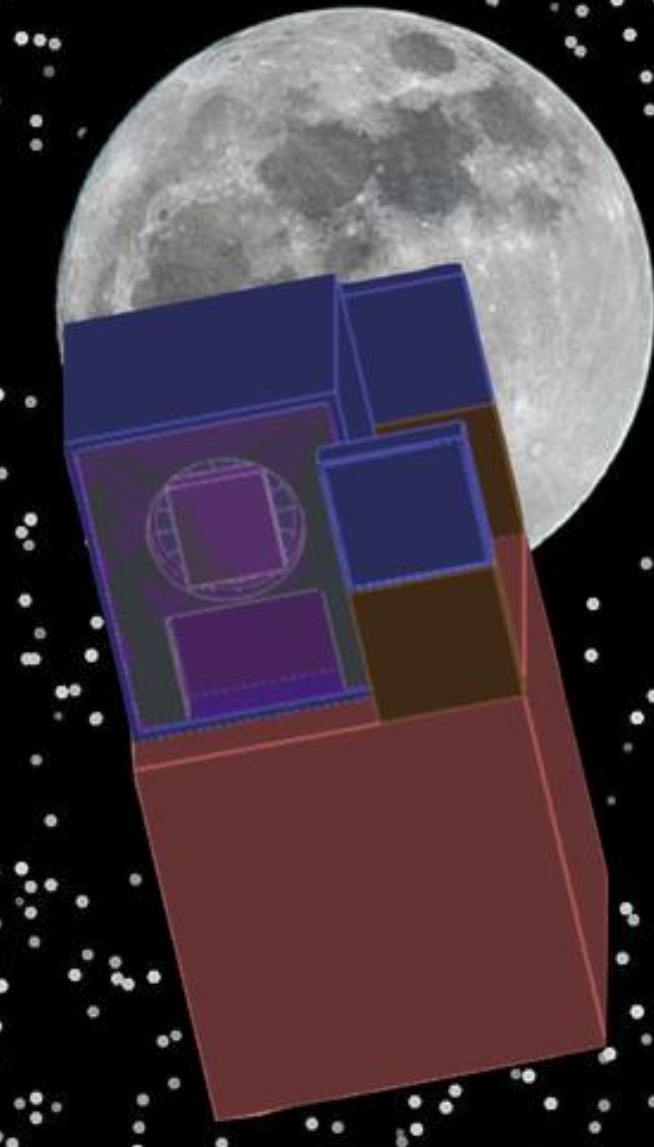}
         \caption{}
         \label{fig:lunpan_geant4}
     \end{subfigure}
     \hfill
     \begin{subfigure}[b]{0.45\textwidth}
         \centering
         \includegraphics[width=0.8\textwidth]{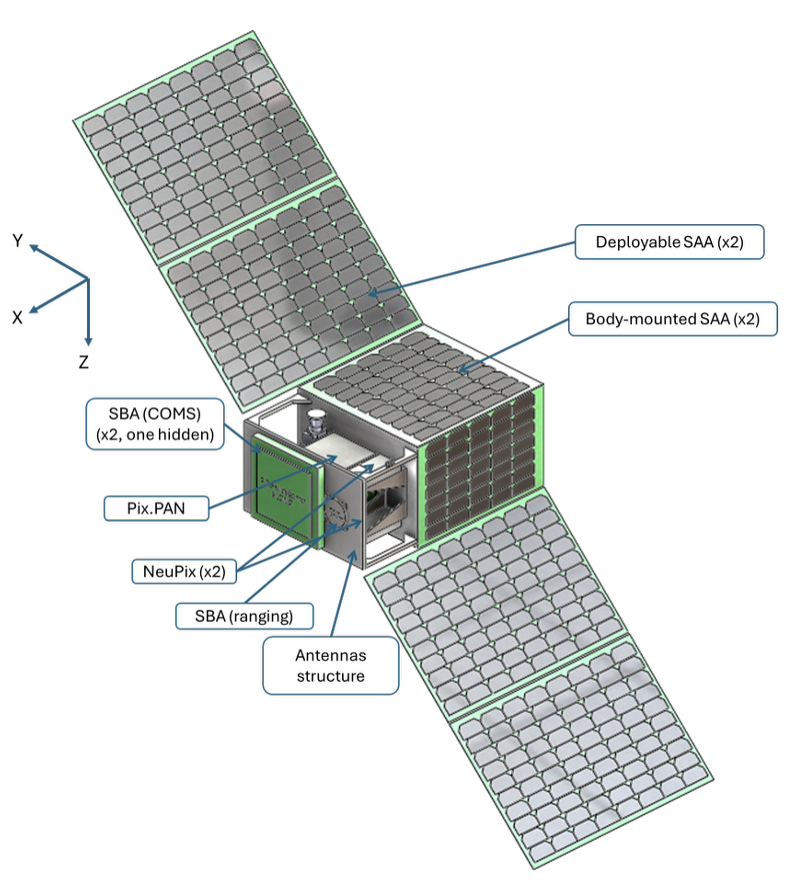}
         \caption{}
         \label{fig:lunpan_platform}
     \end{subfigure}
     \caption{\textbf{a)} Geant4 implementation of LunPAN with its PixPAN (left) and NeuPix payloads (purple and brown squares to the right). \textbf{b)} CAD of the LunPAN platform by AerospaceLab with the 2 payloads}
     \label{fig:lunpan_overview}
\end{figure}

LunPAN (Lunar Particle Analyzer) is a proposed lunar orbiter mission designed to conduct a comprehensive investigation of the energetic particle environment in lunar orbit. The baseline LunPAN mission is planned for a 3-year mission (with a tentative launch in 2030) and to orbit at low lunar orbit (LLO) with an altitude of 100km. Its mission aims to contribute to the fields of galactic cosmic rays (GCRs), solar energetic particles (SEPs), lunar geology, space weather, radiation dosimetry and space technology development.  It does so by providing the flux and composition of penetrating charged particles (100 MeV-10 GeV) with its main payload PixPAN, and energetic charged particles of 10 -100 MeV, as well as neutrons and gammas, with the smaller payload NeuPix (as illustrated in Figure \ref{fig:lunpan_geant4}). By filling the existing measurement gaps in the deep space radiation environment, LunPAN is positioned to make significant contributions to studies on acceleration and propagation processes of GCRs and SEPs, advancing space weather models by providing critical in situ measurements outside the Earth’s magnetosphere, and improving radiation environment models essential for the planning of long-duration lunar missions. Additionally, by measuring albedo neutron and gamma-ray fluxes, LunPAN offers secondary contributions to lunar surface composition studies, including probing hydrogen abundance and elemental distribution, which are relevant for lunar geology and resource exploration.

Figure \ref{fig:lunpan_platform} shows a CAD version of the microsatellite. It is a modified version of the VSP-50 by AerospaceLab, featuring a payload field of view (FoV) that provides an unobstructed view towards nadir (+Z) and zenith (-Z). The communications and ranging subsystem includes two directional S-band antennas for communications and one non-directional antenna for ranging, along with two associated electronic units. When stowed, the platform measures 755x500x390 mm, has a total wet mass of 70.5kg and can ensure a 100\% duty cycle non-Nadir pointing operations. The downlink data rate depends on the final trajectory and the use of the ESA's Moonlight service (up to 4.5GB/day). 

\section{Payload Design}
\subsection{PixPAN} \label{sec:pixpan}

\begin{figure}[hb]
     \centering
     \begin{subfigure}[b]{0.35\textwidth}
         \centering
         \includegraphics[width=\textwidth]{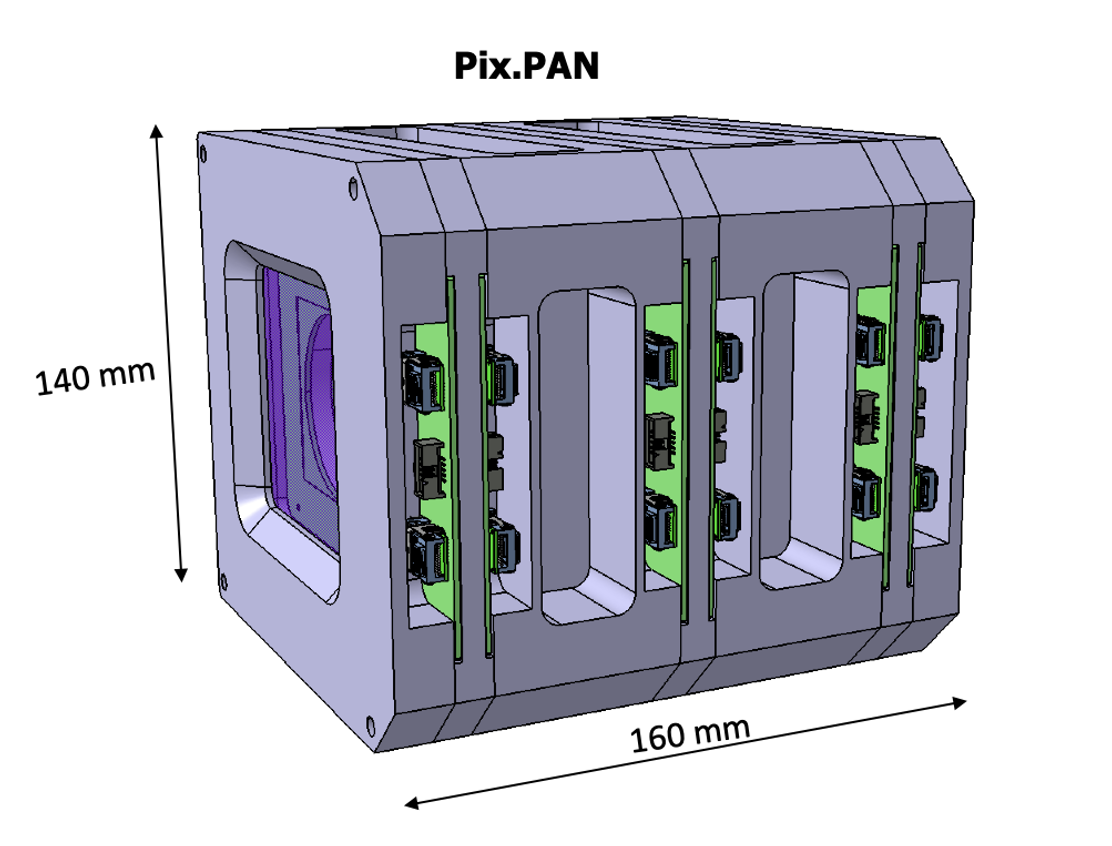}
         \caption{}
         \label{fig:pixpan_cad}
     \end{subfigure}
     \begin{subfigure}[b]{0.3\textwidth}
         \centering
         \includegraphics[width=\textwidth]{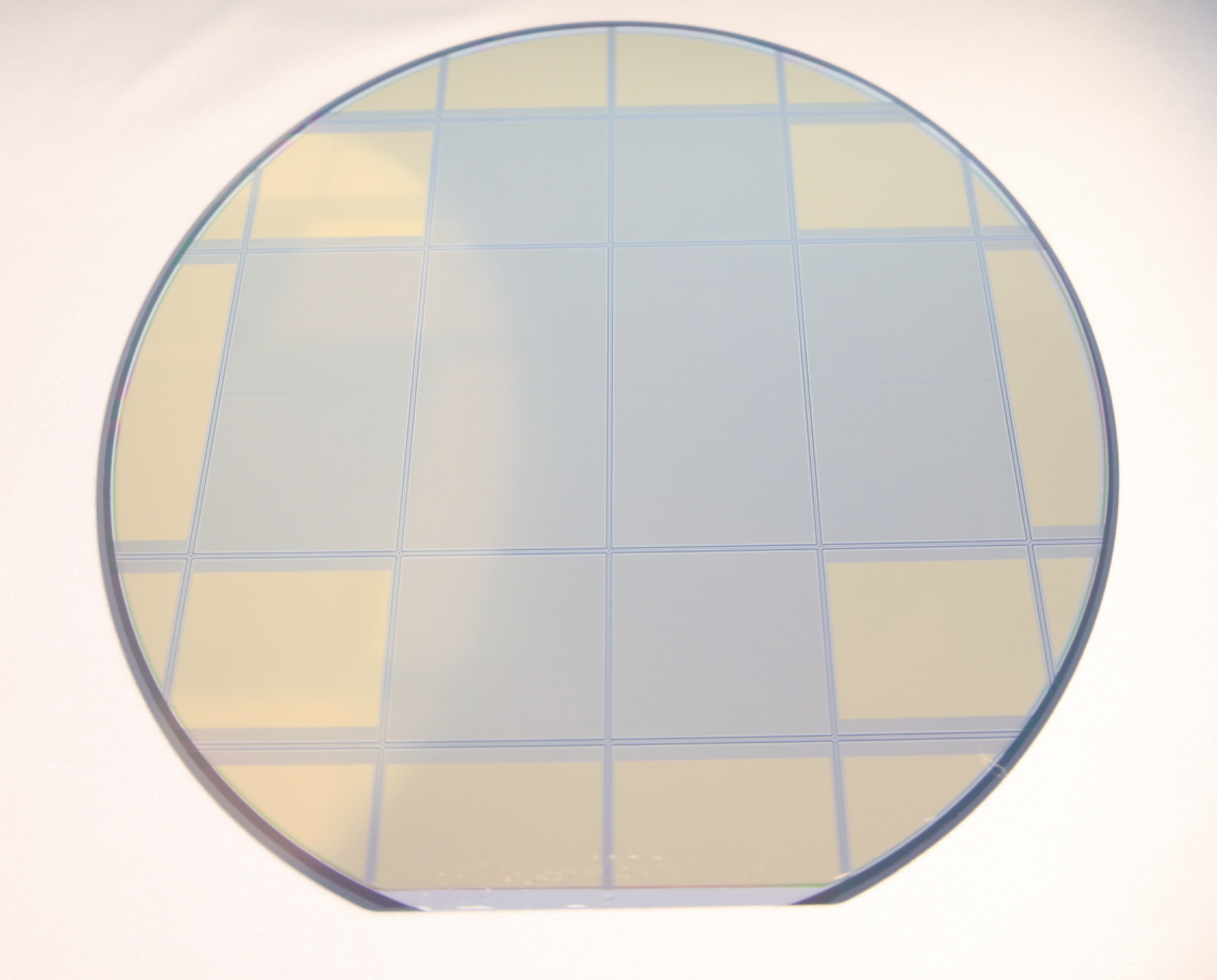}
         \caption{}
         \label{fig:pixpan_wafer}
     \end{subfigure}
     \hfill
     \begin{subfigure}[b]{0.3\textwidth}
         \centering
         \includegraphics[width=\textwidth]{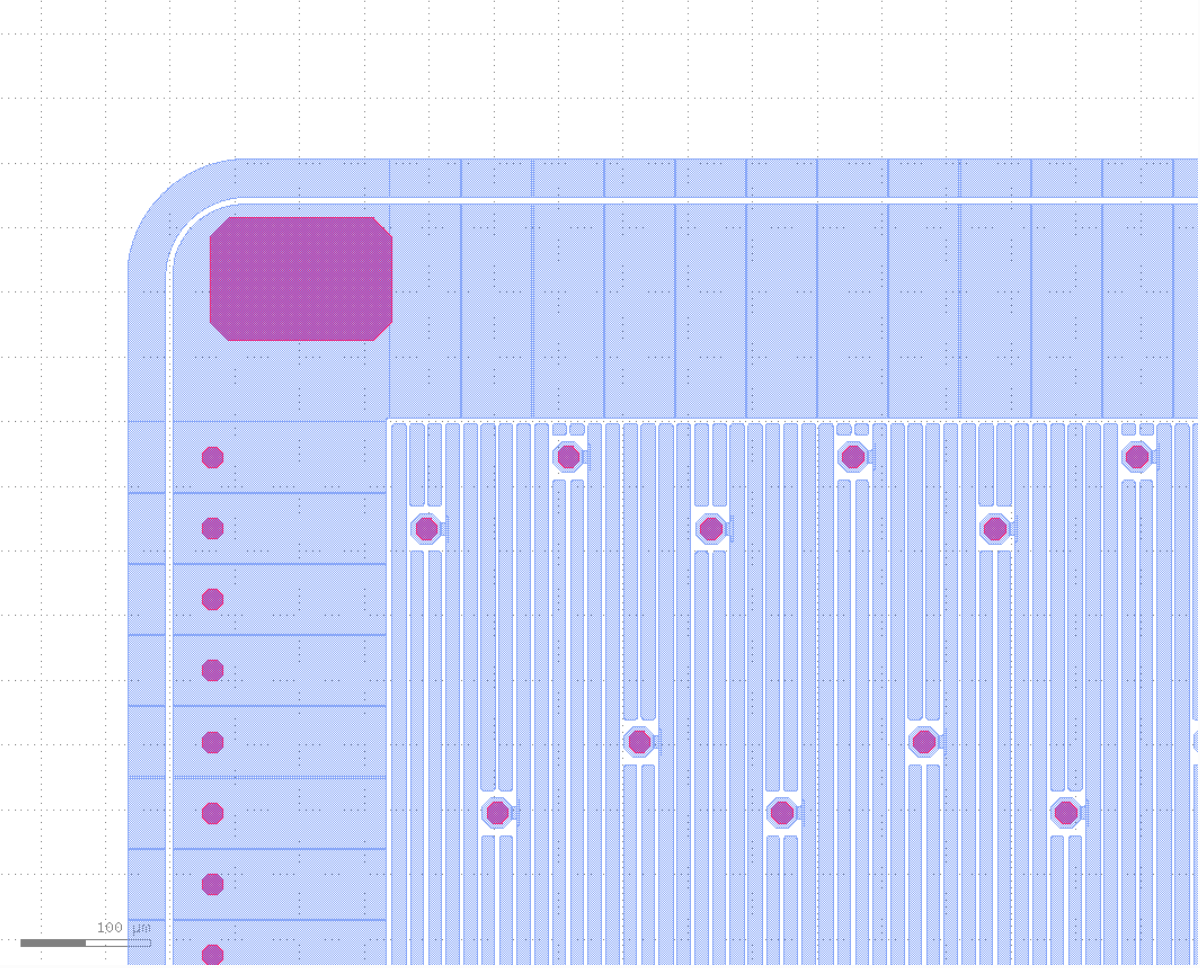}
         \caption{}
         \label{fig:pixpan_sensor_layout}
     \end{subfigure}     
     \caption{\textbf{a)} Preliminary design of the mechanical structure of the PixPAN spectrometer. \textbf{b)} 300$\mu m$ thick Si wafer with "long thin pixels" at Advafab. \textbf{c)} Schematic of the sensor layout and how the "long thin pixels" are connected to the standard Timepix4 square readout cells.}
     \label{fig:pixpan_design}
\end{figure}

Pix.PAN is a compact magnetic spectrometer designed for precise measurements of penetrating charged particles, ranging from 100 MeV to 10 GeV. It builds on the Particle Analyzer (PAN) concept \cite{Wu2019} and the EU H2020 funded Mini.PAN project\cite{Sukhonos2023}. It has also completed the Phase0/A/B1 studies with ESA's "Ambitious Project for the Czech Republic" with the REMEC proposal \cite{BOLIS202565}, and completed the pre-Phase A studies with the COMPASS proposal for a NASA call \cite{Hulsman2023}. Magnetic spectrometers measure the rigidity of a charged particle through its bending in the magnetic field, and therefore derive the momentum and energy if the charge (Z) of the particle can be identified. The baseline layout of PixPAN is shown in Figure \ref{fig:pixpan_design}. It is a cylindrical magnetic spectrometer with two Halbach-array magnet sectors of 5 cm in diameter, each providing a dipole field of 0.4 Tesla. The magnets are sandwiched between three tracking stations, each composed of two tracking layers 1 cm apart, to measure both the particle bending angle through each magnet, as well as its bending radius through the full spectrometer. The overall dimension of the spectrometer is within an envelope of 16×14×14 cm$^3$, with a total weight around 10 kg (margin excluded), and a power consumption around 20 W (excluding data handling). This design has been largely been validated with the Mini.PAN demonstrator \cite{Sukhonos2023} (which was composed of 3 silicon strip trackers and 2 Timepix3 silicon pixel modules). PixPAN will use the same magnets as Mini.PAN and use 3 Timepix4 \cite{Llopart2022} silicon pixels modules, thus making its design simpler and more robust.

Timepix is a series of hybrid pixel detector (HPD) readout ASIC developed by the Medipix2 collaboration led by CERN, and has been widely used in particle physics, nuclear physics, medical physics, and in space. Timepix4 \cite{Llopart2022} is the 4th generation of the ASIC produced and validated in 2021. The PixPAN tracking layer consists of silicon pixel detector readout by 2x2 Timepix4. The Timepix4 chip measures 29.96x24.7 mm$^2$, therefore the quad assembly can read out a silicon pixel detector up to 60x50 mm$^2$ in size, sufficient to cover the sensitive area of the PixPAN spectrometer. Note, this sensor layout has already been successfully produced with Timepix3 for the MiniPAN demonstrator. To achieve the required position resolution for PixPAN with a power budget suitable for a space instrument, and within the current manufacturing capabilities, a sensor with "long thin pixels" of 13.75$\mu m$x1746$\mu m$, mapping to the 55$\mu m$x55$\mu m$ square readout cells with an integrated "pitch adapter" \ref{fig:pixpan_sensor_layout} (developed together with the Finnish company Advafab \ref{fig:pixpan_wafer}). A prototype sensor has been produced and tested with particles at CERN in October 2023. With this sensor a position resolution of 3-5$\mu m$ can be achieved, while the power consumption of the spectrometer can be kept to around 20 W. The energy resolution of PixPAN with sensor thicknesses between 200 and 400$\mu m$ has be simulated for electrons and protons, using the Geant4 simulation \cite{Agostinelli2003}. The expected energy resolution is <12\% for electrons with energies from a few MeV to a few GeV, and of <40\% for protons of $~$100 MeV to a few GeV \cite{Hulsman2023}, fulfilling LunPAN's mission requirements.

\subsection{NeuPix} \label{sec:neupix}

NeuPix is based on the same hybrid pixel dector technology as PixPAN, implemented with the identical Timepix4 readout ASIC, thus greatly reducing the complexity of the system. But a sensor and converter scheme optimized for the detection of neutrons, gamma-rays and 10 MeV - 100 MeV charged particles, already developed for ground-based experiment, will be adopted. NeuPix, is a symmetric instrument same as PixPAN, with one end point to the Moon, and the other to the zenith. This configuration allows to extract the fluxes originated from the Moon, and those from GCR, SEP, in the case of charged particle, and from spacecraft activation in the case of $\gamma$-rays. Each half of the NeuPix consists of a stack of 2 units, the top one for neutron and low energy particle, and the bottom one for $\gamma$-rays. Therefore, together they provide a powerful way to cleanly identify charged particles, neutrons and gamma-rays.

\subsubsection{Neutron Unit}

\begin{figure}[ht]
     \centering
     \begin{subfigure}[b]{0.25\textwidth}
         \centering
         \includegraphics[width=\textwidth]{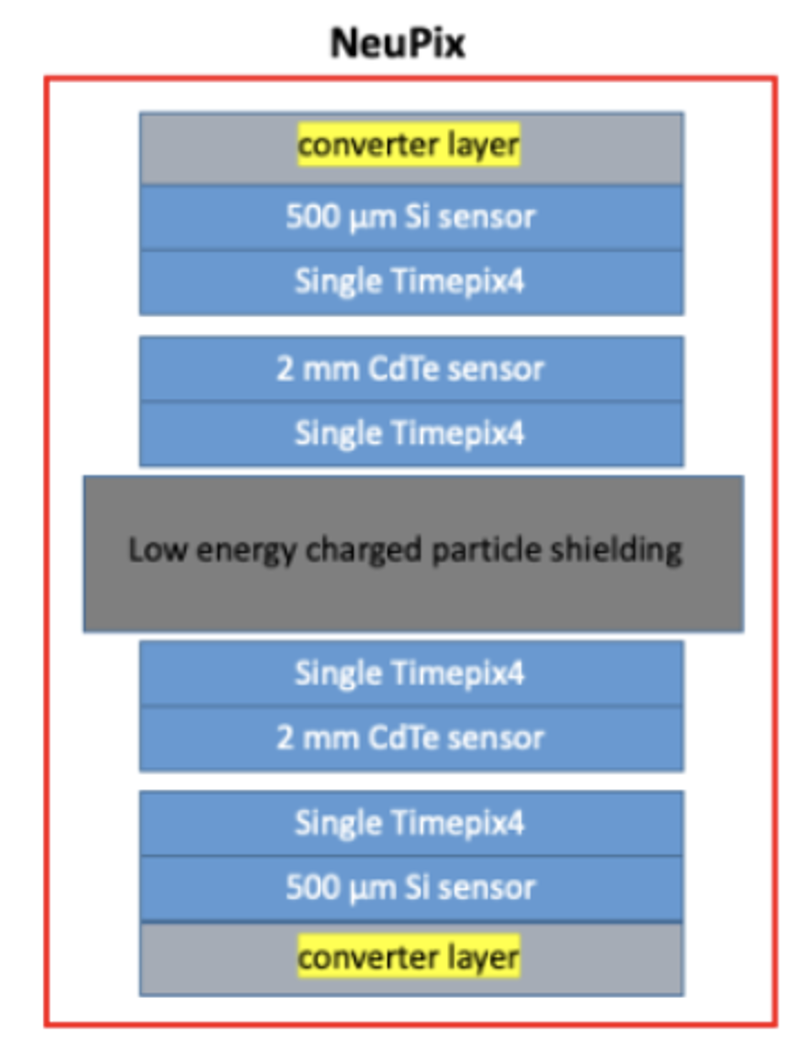}
         \caption{}
         \label{fig:neupix_neutron_1}
     \end{subfigure}
     \hfill
     \begin{subfigure}[b]{0.35\textwidth}
         \centering
         \includegraphics[width=\textwidth]{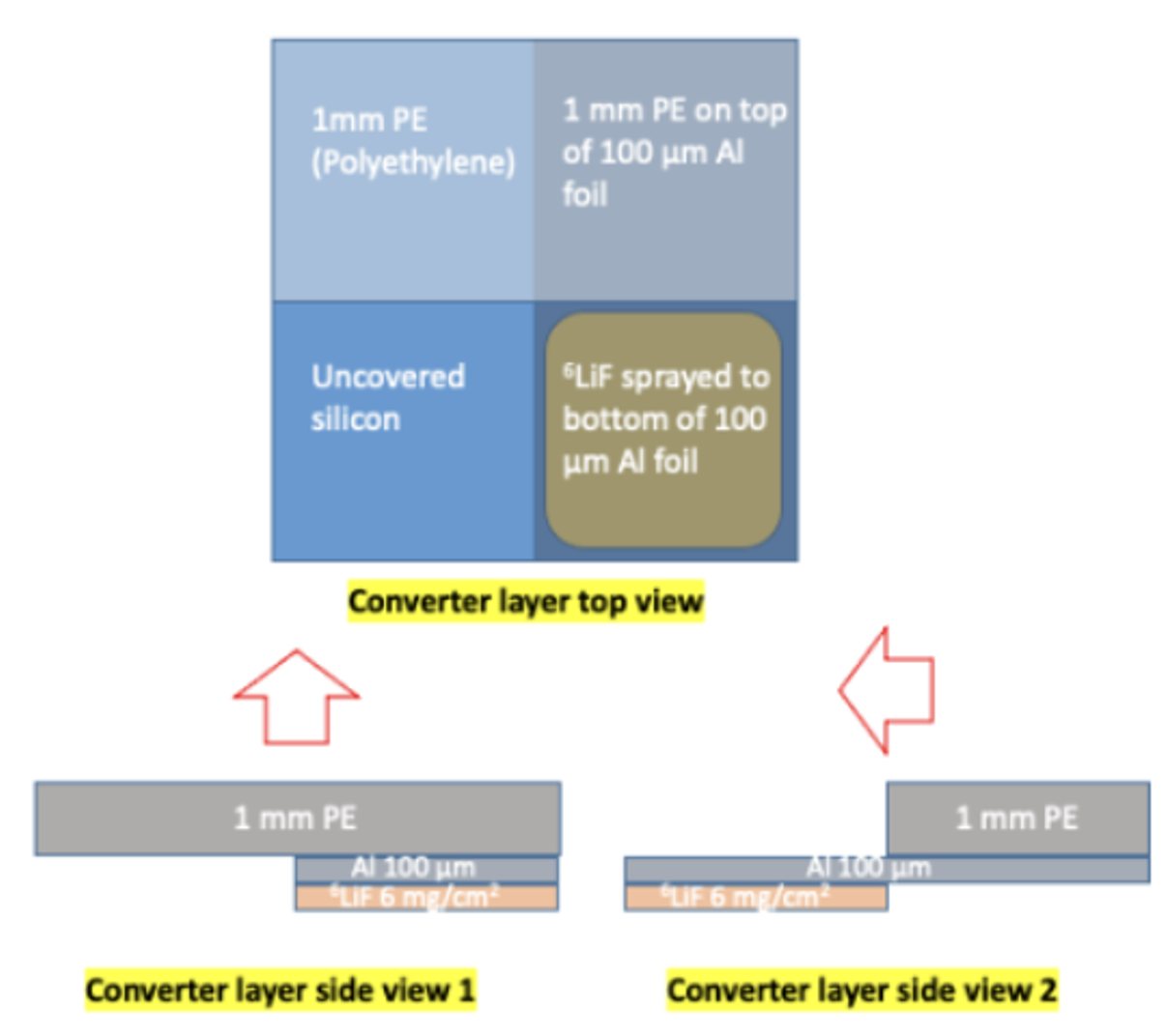}
         \caption{}
         \label{fig:neupix_neutron_2}
     \end{subfigure}
     \begin{subfigure}[b]{0.35\textwidth}
         \centering
         \includegraphics[width=\textwidth]{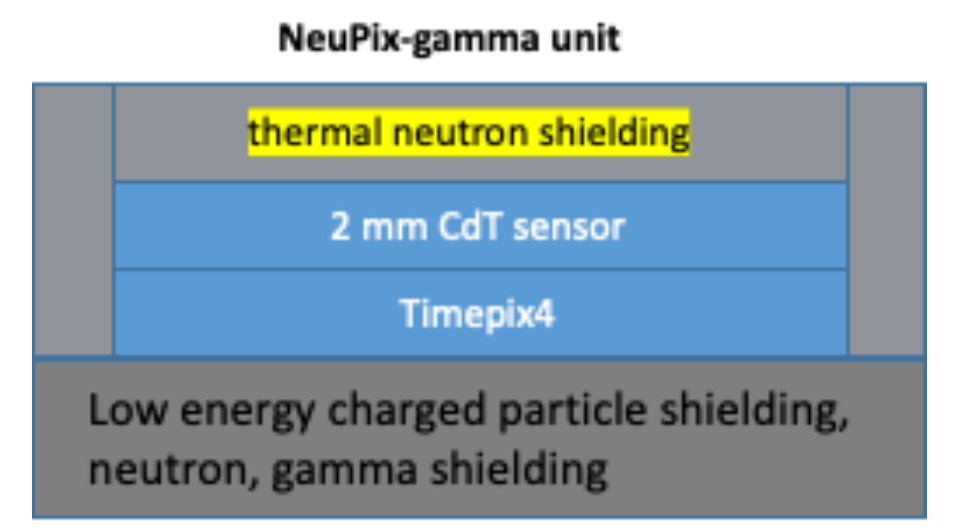}
         \caption{}
         \label{fig:neupix_gamma}
     \end{subfigure}
     \caption{\textbf{a)} Sketch of the NeuPix instrument showing its main components (not to scale). \textbf{b)} Sketch of top view and 2 side views of the converter layer (not to scale) of the \emph{Neutron} unit. \textbf{c)} Sketch of the \emph{Gamma} unit with its shielding and sensor layer.}
     \label{fig:neupix}
\end{figure}

The neutron unit consists of a converter layer, followed by a 500$\mu m$ thick and standard 55$\mu m$ pixel size silicon pixel sensor readout by one TimePix4 ASIC, corresponding to a sensitive area of 2.8 cm x 2.5 cm$^2$. The converter layer of the neutron units has 4 equal-sized segments, as shown in Figure \ref{fig:neupix_neutron_1} and \ref{fig:neupix_neutron_2}:
\begin{itemize}[noitemsep, topsep=0pt]
    \item{\textbf{Low energy segment:}} no converter material, for low energy (10 MeV - 100 MeV) charged particle detection
    \item{\textbf{Thermal neutron segment:}} a 100$\mu m$ Aluminum foil, coated with 6 mg/cm$^2$ Li$^6$F on the side facing the Si sensor, for thermal neutron (0.015 eV - $~$100 keV) detection through neutron capture
    \item{\textbf{Fast neutron segment 1:}} a 1 mm thick polyethylene (plastic) layer, for fast neutron (> 1MeV) detection through n,p scattering
    \item{\textbf{Fast neutron segment 2:}} a 1 mm thick polyethylene (plastic) layer with a 100$\mu m$ Aluminum foil attached at the bottom to allow only higher energy recoil proton to reach the Si sensor, so it is only sensitive to neutrons > 5 MeV
\end{itemize}

\subsubsection{Gamma Unit}
The gamma unit sits right underneath the neutron unit and consists of a 2 mm thick CdTe pixel sensor with standard 55$\mu m$ pixels, readout by the TimePix4 ASIC. Thanks to its fine granularity the gamma unit can distinguish protons, ions and gamma-rays. Therefore, the gamma unit and the neutron unit together provide a power way to cleanly identify charged particles, neutrons and gamma-rays.  A schematic of the cross section of the $\gamma$-unit is shown in Figure \ref{fig:neupix_gamma}.

\section{Data Processing}
\subsection{PixPAN}

Each hit in the Timepix4 detectors is written in 64-bit data packages. The information includes timing, energy and associated housekeeping data. To estimate expected data rate, a closely resembling instrument was implemented in Geant4 (see Figure \ref{fig:lunpan_geant4}; minute details regarding the design were ignored due to the nature of this study). A 5mm Aluminium shield encasing is included. 
The particle spectra are grouped into GCR, lunar albedo and solar flare particles. In the estimation we account for impact of the shadow from the moon with respect to its distance. For the solar flare particles we use the Oct. 1989 Solar Flare spectrum (obtained from the SPENVIS framework \cite{Heynderickx2004}). For this analysis and to maximally obtain all data for the science case, the threshold is actually set to be as low as possible; 1000e- (or about 3.5 keV). With this threshold setting, the total expected data rate is $~$35.2kB/s (100km) and $~$43.4kB/s (1000km) as derived from Table \ref{tab:pixpan_data_rates}. It should be noted however that this rate reflects the raw data rate (i.e., every hit). This may be useful for commissioning purposes or similar, but the particles must pass through multiple layers to be considered useful for science analysis. A simple on-board time-of-arrival comparison would suffice as a first level event selection criterion. If the particles must pass through 1 magnet segment (i.e., 4 tracker layers – Mode 1) or through 2 magnet segments (i.e., 6 tracker layers – Mode 2), the data rate will drop as depicted in the table. The same analysis is performed for the Oct. 1989 Solar Flare. Assuming the same energy threshold we expect the data rate to be $~$8700 kB/s, $~$852 kB/s and $~$251 kB/s for the raw, Mode 1 and Mode 2 data rates respectively. This analysis therefore shows that to comply with the data rate limit, a first-level on-board event selection (based on timing) is necessary. As solar flares are temporary events, a hard drive would suffice to act as a "buffer" to store the data while transmitting the information to Earth. The solar flare considered here, at its peak, reached a flux of more than $10^5$ counts/cm$^2$/sec/sr for >10 MeV protons (i.e. an S5 class). These events occur on average less than 1 per solar cycle. For $~$3 years strong solar flares (S3 class) can be used as a representable reference. For an S3 class flare the flux level is two orders of magnitude lower compared to the S5. We therefrom find that a 10 GB hard drive would suffice for the mission. 

\begin{table}[h!]
\centering
\setlength{\tabcolsep}{8pt}

\rowcolors{3}{white}{white} 

\begin{tabular}{>{\bfseries}l
                >{\centering\arraybackslash}p{2cm}
                >{\centering\arraybackslash}p{2cm}
                >{\centering\arraybackslash}p{2cm}
                >{\centering\arraybackslash}p{2cm}}
\toprule
\rowcolor[HTML]{E5E5E5} 
 & \multicolumn{2}{c}{\textbf{GCR Rate in kB/s}} & \multicolumn{2}{c}{\textbf{Albedo Rate in kB/s}} \\
\cmidrule(lr){2-3} \cmidrule(lr){4-5}
 & \textbf{Data Rate @ 100km} & \textbf{Data Rate @ 1000km} & \textbf{Data Rate @ 100km} & \textbf{Data Rate @ 1000km} \\
\midrule
\textbf{PixPAN (raw)} & 28.859 & 40.483 & 6.314 & 2.934 \\
\textbf{PixPAN (Mode 1)}   & 2.679 & 3.971  & 0.548 & 0.249 \\
\textbf{PixPAN (Mode 2)}   & 0.485 & 0.766  & 0.068 & 0.076 \\
\bottomrule
\end{tabular}
\caption{PixPAN data rate for GCR and albedo particles with respect to distance to the lunar surface}
\label{tab:pixpan_data_rates}
\end{table}

Subsequent science analyses can be executed in post upon transmission to Earth. Each Timepix4 quad will need to be time and energy calibrated. This helps determine the actual time stamp and energy deposit on an instrument level. It is also important to consider the alignment of the layers. Small manufacturing errors and/or shifts due to mechanical vibrations may cause minute shifts (on $\mu m$ level) between each tracker level. Without accounting for this shift the reconstruction algorithm will be worse. As particle types interact differently with the silicon sensor, their cluster structure (neighbouring hits from same incident particle) will also look differently. The cluster algorithm finds these clusters which can then be used to perform the final particle reconstruction. Finally, after all these steps/considerations, the scientific results will be at the desired quality for publishing.

\subsection{NeuPix}

NeuPix units will process data onboard and produce particle species identification, fluxes, dose rates and energy spectroscopy for each particle type based on the morphology of each detected particle track. These are final science products and no further processing on ground is expected. Occasional raw data will be taken and transferred to check the status of sensors and to verify the correct performance of the onboard processing. Convolutional neural network are used to process these data and to obtain particle species identification, fluxes, dose rates, energy spectroscopy and directional information for each particle species. Considering that the detector will be receiving $~$350 particles or 5000 hits from the space environment every second, it means NeuPix will be collecting 36 kB/s or 1.1 GB/day worth of space environment events. To reduce data produced and to be sent to the ground, 3 modes are developed:
\begin{itemize}[noitemsep, topsep=0pt]
    \item{\textbf{Raw mode:}} This mode sends all events recorded in the NeuPix instrument; data rate of 96 kB/s. (used during short periods for on-ground fine tuning)
    \item{\textbf{List mode:}} This mode sends a list of cluster parameters (angles, energy, size, linearity, ...); data rate roughly constant at 21.6 kB/s. (helpful during SPEs by having a 20 or more-compression factor)
    \item{\textbf{Nominal/Histogram mode:}} This mode sends 7 histograms every day for the electron, proton, thermal neutron flux, fast neutron flux > 1 MeV, fast neutron > 4 MeV, $\gamma-$-rays and dE/dX Proton spectrum, data rate fixed of 53 kB/day (<1B/s).
\end{itemize}

NeuPix nominal mode will be the Histogram mode allowing to have a low data rate of 53 kB/day. During given period, NeuPix should be switched to the Raw and List mode for total data transparency and access to primary events which will generate 36 kB/s and 21.6 kB/s respectively. However, Raw and List mode depends on the space weather condition, as more events are generated when NeuPix receives more particles. A rough estimate of the Raw data rate of NeuPix is given in Table \ref{tab:neupix_data_rates}, this estimate was calculated with real simulation of the space environment and NeuPix geometry factor. 

\begin{table}[h!]
\centering
\setlength{\tabcolsep}{8pt}

\rowcolors{3}{white}{white} % reset default alternating colors

\begin{tabular}{>{\bfseries}l
                >{\centering\arraybackslash}p{2cm}
                >{\centering\arraybackslash}p{2cm}
                >{\centering\arraybackslash}p{2cm}
                >{\centering\arraybackslash}p{2cm}}
\toprule
\rowcolor[HTML]{E5E5E5} 
 & \multicolumn{2}{c}{\textbf{GCR Rate in kB/s}} & \multicolumn{2}{c}{\textbf{Albedo Rate in kB/s}} \\
\cmidrule(lr){2-3} \cmidrule(lr){4-5}
 & \textbf{Data Rate @ 100km} & \textbf{Data Rate @ 1000km} & \textbf{Data Rate @ 100km} & \textbf{Data Rate @ 1000km} \\
\midrule
\textbf{NeuPix (raw; Neutron)} & 8.52 & 12.18 & 29.55 & 14.65 \\
\textbf{NeuPix (raw; Gamma)}   & 3.60 & 5.22  & 12.58 & 6.55 \\
\bottomrule
\end{tabular}
\caption{NeuPix data rate for GCR and albedo particles with respect to distance to the lunar surface}
\label{tab:neupix_data_rates}
\end{table}

\section{Orbit Selection}

\begin{figure}[ht!]
     \centering
     \begin{subfigure}[b]{0.49\textwidth}
         \centering
         \includegraphics[width=\textwidth]{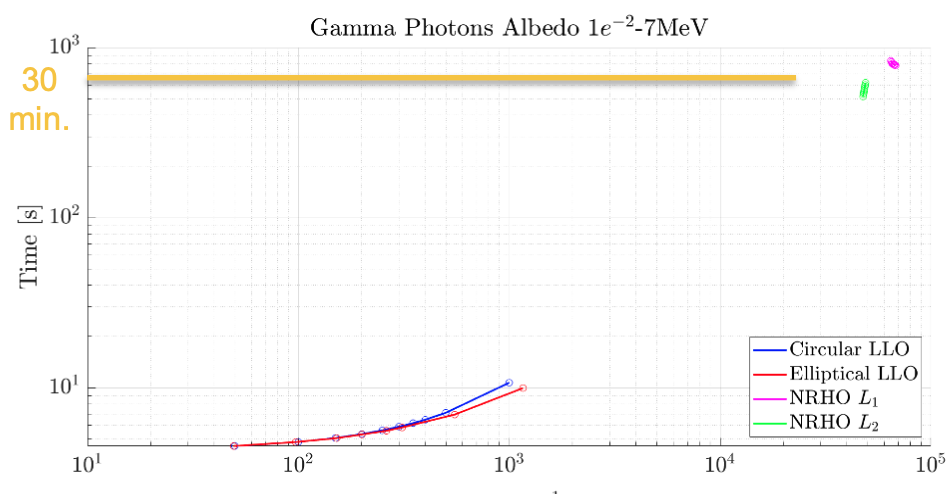}
         \caption{}
         \label{fig:gamma_rate}
     \end{subfigure}
     \hfill
     \begin{subfigure}[b]{0.49\textwidth}
         \centering
         \includegraphics[width=\textwidth]{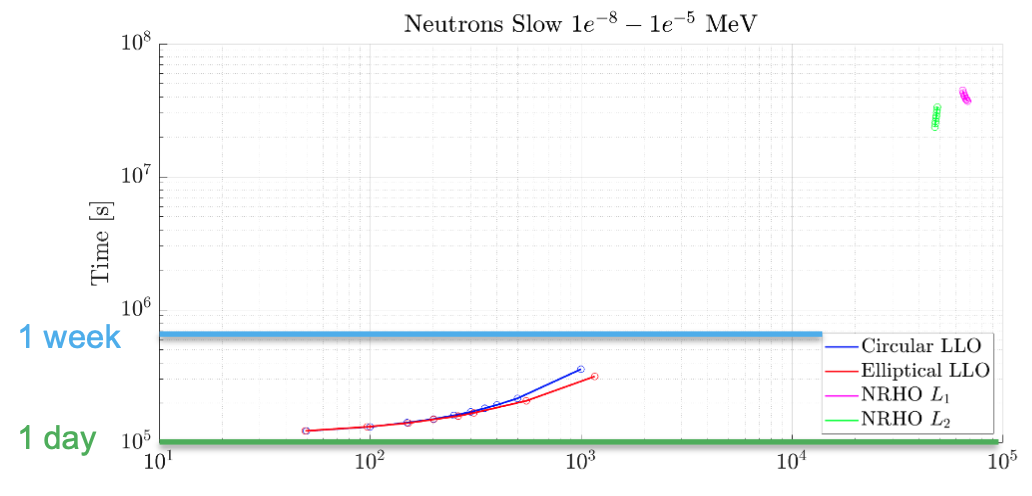}
         \caption{}
         \label{fig:slow_neutrons_rate}
     \end{subfigure}
     \begin{subfigure}[b]{0.49\textwidth}
         \centering
         \includegraphics[width=\textwidth]{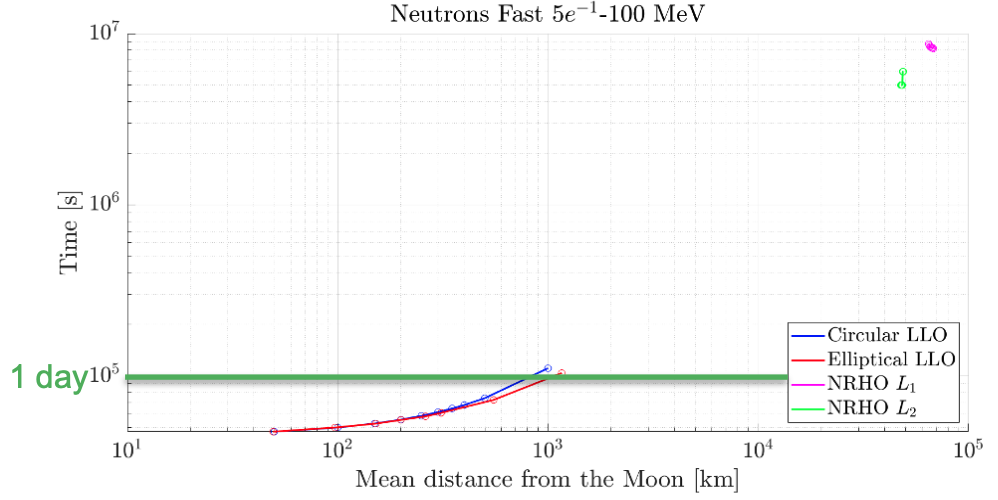}
         \caption{}
         \label{fig:fast_neutron_rate}
     \end{subfigure}
     \hfill
     \begin{subfigure}[b]{0.49\textwidth}
         \centering
         \includegraphics[width=\textwidth]{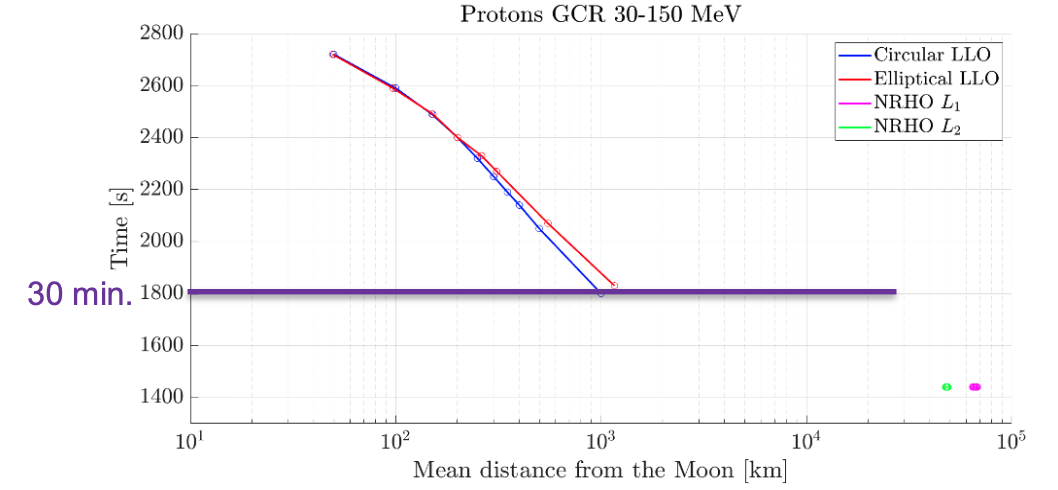}
         \caption{}
         \label{fig:proton_GCR_rate}
     \end{subfigure}
     \caption{Estimated times needed to reach 5\% Poissonian uncertainty for \textbf{a)} albedo $\gamma$,  \textbf{b)} slow neutrons,  \textbf{c)} fast neutrons and \textbf{d)} GCR protons. Estimated times for LLOs and NRHOs are compared in each figure.}
     \label{fig:particle_rate}
\end{figure}

For LunPAN, the best choice of orbit is where the total number of albedo particles (in particular for neutrons) is maximal. A trajectory where the spacecraft is far from the lunar surface for a significant amount of time will result in lower particle counts and should be avoided. Also, due to the decay time of neutron, orbits with a large apogee result in a cut-off of neutrons (distance depends on momentum of neutron). Regarding the SEP and GCR, the orbit selection has little significant impact, resulting from the higher particle flux.
A custom tool was created to help compare between various proposed orbits, calculating the average particle flux on a given orbit and returns the time needed to detect 400 particles (5\% Poissonian uncertainty) of a given type and energy range. As the Figures in \ref{fig:particle_rate} show, approximately 2 days on a low lunar orbit equates to around 1 year on a NRHO orbit to detect 400 slow neutrons (Figure \ref{fig:slow_neutrons_rate}). For other particle species less time is needed. Therefore, the science goals are addressed best when selecting the LLO. As part of the pre-A studies, a stable orbit was found for the duration of LunPAN's mission. THe baseline orbit solution is a dedicated launch with a direct insertion of the LLO by the launcher. The estimated station-keeping cost is expected to be 150m/s/year (to be consolidated in a later stage).

\section{Summary}

LunPAN is a novel mission proposal, designed to provide a comprehensive characterization of the lunar energetic particle environment through its synergistic PixPAN and NeuPix instruments, addressing key observation gaps in low (10–100 MeV) and high (100 MeV–10 GeV) energy particle measurements in deep space, providing valuable data for the Galactic Cosmic Rays, Solar Energetic Particles, and lunar albedo particles science fields. As part of the pre-A mission studies, a platform by AerospaceLab was identified with a closed mass budget of 70.5 kg, providing enough power for a 100\% duty cycle in non-nadir pointing operations. A stable 100km LLO orbit was also identified, allowing the payload to acquire the 5\% statistical level within 2 days for all relevant particles species (compared to $~$1 year for a NRHO). 

\bibliography{references}

@inproceedings{Sukhonos2023,
  series = {ICRC2023},
  title = {Penetrating particle ANalyzer (PAN)},
  url = {http://dx.doi.org/10.22323/1.444.0045},
  DOI = {10.22323/1.444.0045},
  booktitle = {Proceedings of 38th International Cosmic Ray Conference — PoS(ICRC2023)},
  publisher = {Sissa Medialab},
  author = {Sukhonos,  Daniil and Ambrosi,  Giovanni and Azzarello,  Philipp and Barbanera,  Mattia and Bergmann,  Benedikt and Burian,  Petr and Cadoux,  Franck and Favre,  Yannick and Hulsman,  Johannes and Iizawa,  Tomoya and Ionica,  Maria and Marra,  Daniel La and Mancini,  Eduardo and Nicola,  Laurent and Paniccia,  Mercedes and Silvestre,  Gianluigi and Smolyanskiy,  Petr and Stauffer,  Jér\^ome and Stil,  Adrien and Thonet,  Pierre Alexandre and Xie,  Pengwei and Wu,  Xin},
  year = {2023},
  month = aug,
  pages = {045},
  collection = {ICRC2023}
}

@article{BOLIS202565,
title = {Mission analysis for the Radiation Environment Monitor for Energetic Cosmic rays (REMEC) mission},
journal = {Acta Astronautica},
volume = {230},
pages = {65-78},
year = {2025},
issn = {0094-5765},
doi = {https://doi.org/10.1016/j.actaastro.2025.02.013},
url = {https://www.sciencedirect.com/science/article/pii/S0094576525000840},
author = {Mathilda Bolis and Elisa Maria Alessi and Camilla Colombo}
}

@article{Hulsman2023,
  title = {Relativistic particle measurement in jupiter’s magnetosphere with Pix.PAN},
  volume = {56},
  ISSN = {1572-9508},
  url = {http://dx.doi.org/10.1007/s10686-023-09918-4},
  DOI = {10.1007/s10686-023-09918-4},
  number = {2–3},
  journal = {Experimental Astronomy},
  publisher = {Springer Science and Business Media LLC},
  author = {Hulsman,  Johannes and Wu,  Xin and Azzarello,  Philipp and Bergmann,  Benedikt and Campbell,  Michael and Clark,  George and Cadoux,  Franck and Ilzawa,  Tomoya and Kollmann,  Peter and Llopart,  Xavi and Nénon,  Quentin and Paniccia,  Mercedes and Roussos,  Elias and Smolyanskiy,  Petr and Sukhonos,  Daniil and Thonet,  Pierre Alexandre},
  year = {2023},
  month = nov,
  pages = {371–402}
}

@article{Wu2019,
  title = {Penetrating particle ANalyzer (PAN)},
  volume = {63},
  ISSN = {0273-1177},
  url = {http://dx.doi.org/10.1016/j.asr.2019.01.012},
  DOI = {10.1016/j.asr.2019.01.012},
  number = {8},
  journal = {Advances in Space Research},
  publisher = {Elsevier BV},
  author = {Wu,  X. and Ambrosi,  G. and Azzarello,  P. and Bergmann,  B. and Bertucci,  B. and Cadoux,  F. and Campbell,  M. and Duranti,  M. and Ionica,  M. and Kole,  M. and Krucker,  S. and Maehlum,  G. and Meier,  D. and Paniccia,  M. and Pinsky,  L. and Plainaki,  C. and Pospisil,  S. and Stein,  T. and Thonet,  P.A. and Tomassetti,  N. and Tykhonov,  A.},
  year = {2019},
  month = apr,
  pages = {2672–2682}
}

@article{Llopart2022,
  title = {Timepix4,  a large area pixel detector readout chip which can be tiled on 4 sides providing sub-200 ps timestamp binning},
  volume = {17},
  ISSN = {1748-0221},
  url = {http://dx.doi.org/10.1088/1748-0221/17/01/c01044},
  DOI = {10.1088/1748-0221/17/01/c01044},
  number = {01},
  journal = {Journal of Instrumentation},
  publisher = {IOP Publishing},
  author = {Llopart,  X. and Alozy,  J. and Ballabriga,  R. and Campbell,  M. and Casanova,  R. and Gromov,  V. and Heijne,  E.H.M. and Poikela,  T. and Santin,  E. and Sriskaran,  V. and Tlustos,  L. and Vitkovskiy,  A.},
  year = {2022},
  month = jan,
  pages = {C01044}
}

@article{Agostinelli2003,
  title = {Geant4—a simulation toolkit},
  volume = {506},
  ISSN = {0168-9002},
  url = {http://dx.doi.org/10.1016/S0168-9002(03)01368-8},
  DOI = {10.1016/s0168-9002(03)01368-8},
  number = {3},
  journal = {Nuclear Instruments and Methods in Physics Research Section A: Accelerators,  Spectrometers,  Detectors and Associated Equipment},
  publisher = {Elsevier BV},
  author = {Agostinelli,  S. and Allison,  J. and Amako,  K. and Apostolakis,  J. and Araujo,  H. and Arce,  P. and Asai,  M. and Axen,  D. and Banerjee,  S. and Barrand,  G. and Behner,  F. and Bellagamba,  L. and Boudreau,  J. and Broglia,  L. and Brunengo,  A. and Burkhardt,  H. and Chauvie,  S. and Chuma,  J. and Chytracek,  R. and Cooperman,  G. and Cosmo,  G. and Degtyarenko,  P. and Dell’Acqua,  A. and Depaola,  G. and Dietrich,  D. and Enami,  R. and Feliciello,  A. and Ferguson,  C. and Fesefeldt,  H. and Folger,  G. and Foppiano,  F. and Forti,  A. and Garelli,  S. and Giani,  S. and Giannitrapani,  R. and Gibin,  D. and Gómez Cadenas,  J.J. and González,  I. and Gracia Abril,  G. and Greeniaus,  G. and Greiner,  W. and Grichine,  V. and Grossheim,  A. and Guatelli,  S. and Gumplinger,  P. and Hamatsu,  R. and Hashimoto,  K. and Hasui,  H. and Heikkinen,  A. and Howard,  A. and Ivanchenko,  V. and Johnson,  A. and Jones,  F.W. and Kallenbach,  J. and Kanaya,  N. and Kawabata,  M. and Kawabata,  Y. and Kawaguti,  M. and Kelner,  S. and Kent,  P. and Kimura,  A. and Kodama,  T. and Kokoulin,  R. and Kossov,  M. and Kurashige,  H. and Lamanna,  E. and Lampén,  T. and Lara,  V. and Lefebure,  V. and Lei,  F. and Liendl,  M. and Lockman,  W. and Longo,  F. and Magni,  S. and Maire,  M. and Medernach,  E. and Minamimoto,  K. and Mora de Freitas,  P. and Morita,  Y. and Murakami,  K. and Nagamatu,  M. and Nartallo,  R. and Nieminen,  P. and Nishimura,  T. and Ohtsubo,  K. and Okamura,  M. and O’Neale,  S. and Oohata,  Y. and Paech,  K. and Perl,  J. and Pfeiffer,  A. and Pia,  M.G. and Ranjard,  F. and Rybin,  A. and Sadilov,  S. and Di Salvo,  E. and Santin,  G. and Sasaki,  T. and Savvas,  N. and Sawada,  Y. and Scherer,  S. and Sei,  S. and Sirotenko,  V. and Smith,  D. and Starkov,  N. and Stoecker,  H. and Sulkimo,  J. and Takahata,  M. and Tanaka,  S. and Tcherniaev,  E. and Safai Tehrani,  E. and Tropeano,  M. and Truscott,  P. and Uno,  H. and Urban,  L. and Urban,  P. and Verderi,  M. and Walkden,  A. and Wander,  W. and Weber,  H. and Wellisch,  J.P. and Wenaus,  T. and Williams,  D.C. and Wright,  D. and Yamada,  T. and Yoshida,  H. and Zschiesche,  D.},
  year = {2003},
  month = jul,
  pages = {250–303}
}

@article{Heynderickx2004,
  title = {New radiation environment and effects models in the European Space Agency’s Space Environment Information System (SPENVIS)},
  volume = {2},
  ISSN = {1542-7390},
  url = {http://dx.doi.org/10.1029/2004SW000073},
  DOI = {10.1029/2004sw000073},
  number = {10},
  journal = {Space Weather},
  publisher = {American Geophysical Union (AGU)},
  author = {Heynderickx,  D. and Quaghebeur,  B. and Wera,  J. and Daly,  E. J. and Evans,  H. D. R.},
  year = {2004},
  month = oct 
}
\bibliographystyle{aasjournal}

\end{document}